\documentclass[12pt]{article}
\usepackage{amssymb,amsmath,latexsym,graphicx, bm, mathrsfs, hyperref}
\usepackage{graphics,graphicx}
\usepackage[dvipsnames]{xcolor}

\begin{document}
\newtheorem{prop}{Proposition}
\newtheorem{teo}{Theorem}
\newtheorem{proof}{Proof}

\title{Eisenhart lift and Randers-Finsler formulation \\ for scalar field theory  \vspace{-0.25cm} 
\author{Sumanto Chanda$^1$, Partha Guha$^2$ \\ \vspace{-0.5cm} \\ $ ^1$  
\textit{International Centre for Theoretical Sciences} \\ 
\textit{No. 151, Shivakote, Hesaraghatta Hobli, Bengaluru, Karnataka 560089, India.} 
\\ \texttt{\small sumanto.chanda@icts.res.in} \vspace{0.25cm} \\ 
$ ^2$ \textit{Department of Mathematics, Khalifa University, Zone 1} \\ 
\textit{Abu Dhabi, United Arab Emirates.} \\ 
\texttt{\small partha.guha@ku.ac.ae}} \\
}

\maketitle

\thispagestyle{empty}

\abstract{We study scalar field theory as a generalization of point particle mechanics using 
the Polyakov action, and demonstrate how to extend Lorentzian and Riemannian 
Eisenhart lifts to the theory in a similar manner. Then we explore extension of the 
Randers-Finsler formulation and its principles to the Nambu-Goto action, and 
describe a Jacobi Lagrangian for it.}

\numberwithin{equation}{section}

\section{Introduction}
\label{intro}

The application of geometric methods in physics led to modern theories at the heart 
of active research, such as String Theory and Braneworld Cosmology. The Kaluza-Klein 
theory is one such topic, where one adds extra dimensions to account for other 
interactions,  while the Eisenhart lift is another that formulates a new background curvature that 
replaces the gauge fields that influence a particle's motion by adding degrees of freedom 
associated with symmetry.  The original idea discussed by Eisenhart in 1928 \cite{eisenhart} 
was forgotten with minimal impact until independent rediscovery by Duval in 1985,  
and further developed and studied by Gibbons, Horvathy and others 
\cite{dbkp,dgh,CarAlv,ghkw,CarGib,FilGal,Car,cghhz,GalMas,ForGal,CarGal,chanda}. 
\smallskip

The lifting concept is not completely new, being closely related to Jacobi-Maupertuis 
metric formulation \cite{chanda,cgg1,cggmw}, that does the opposite by 
formulating gauge fields from the curvature to replace the degree of freedom associated 
with symmetry. When a theory is potential free, the corresponding equation is 
tautologically geodesic. Adding potentials to the theory causes a deviation from geodesic. 
However, by using the Eisenhart lift, this can be embedded into the geodesic equation 
of the metric in higher-dimensional space with one extra fictious field. \smallskip

Scalar field theory in n-dimensional field space is a generalization of point particle 
mechanics, which leads us to ask if we can also similarly perform an Eisenhart lift of 
the scalar field space. The main obstacle to performing the Eisenhart lift on field space 
is the divergence term involved in the field equations. Due to this, Finn, Karamitsos 
and Pilaftsis have demonstrated how to execute the Riemannian lift \cite{fkp} of a scalar 
field space with Lagrangian of the Polyakov form by introducing vector fields. \smallskip

The Nambu-Goto action \cite{zwiebach,polchinski} is a geometric way to write the action 
for a string traversing spacetime, similar to the first term in the Randers-Finsler geodesic. 
While other extended objects can be considered for world volume actions, the string alone 
exhibits Weyl invariance. Since it is a geometric action, one can ask if and how the 
formulation and principles associated with Randers-Finsler geodesics discussed in 
\cite{chanda} can be extended to it. \smallskip

The goal of this paper is  to  study  the  Eisenhart  lift  of the scalar field space, and extend 
mechanical principles associated with point particle, world volume actions. We shall thoroughly 
describe both, the Riemannian and Lorentzian lifting processes, in a simple and original way to 
generalize the Eisenhart lift for application on scalar field space. Then the general geometric 
action for a world volume will be considered to study how the Maupertuis principle, constraint, 
a generator analogous to the Hamiltonian, and possibly a Jacobi-Maupertuis formulation may 
be described for it, with the specific case of a surface swept by a string being briefly discussed.

\section{Preliminaries: Scalar Field Theory}
\label{prelim}

While scalar field theory is a generalization of point particle mechanics, 
special circumstances in the latter allow a conserved quantity to exist due to 
an available cyclic co-ordinate, which is not normally possible in the former. 
In this section, we shall review scalar field theory formulation, discuss 
how a conserved quantity can exist if a cyclic field is available, and introduce 
a generator of field equations analogous to the Hamiltonian. \bigskip

\noindent
If we have a $N+1$-dimensional scalar field space $\{ \varphi^i (\bm x) \}$ 
and define the Lagrangian 
$\mathcal L = \mathcal L (g_{\mu \nu}, \varphi^i, \partial_\mu \varphi^i)$ 
on the space, the field indices in configuration space $ 0 \leq i, j, k, m \leq N$, 
co-ordinate indices in base space $0 \leq \mu, \nu \leq n$ and the field action:
\begin{equation} \label{fldact} 
S = \int_{\mathcal V} d^n x \sqrt{- g} \; 
\mathcal L (g_{\mu \nu}, \varphi^i, \partial_\mu \varphi^i), 
\qquad \partial_\mu \varphi^i = \frac{\partial \varphi^i}{\partial x^\mu}
\end{equation}
then remembering that $\delta (\partial_\mu \varphi^i) = 
\partial_\mu (\delta \varphi^i) = \nabla_\mu (\delta \varphi^i)$, the arbitrary 
variation of the action \eqref{fldact} in the spacetime region $\mathcal V$ can 
be expanded using chain rule:
\begin{equation} \label{fldvar2} 
\begin{split}
\delta S &= \int_{\mathcal V} d^n x \sqrt{- g} \left[ 
\left( \frac12 g^{\mu \nu} \mathcal L + 
\frac{\partial \mathcal L}{\partial g_{\mu \nu}} \right) \delta g_{\mu \nu} \right. \\ 
& \qquad \quad  + \left. \left\{ \frac{\partial \mathcal L}{\partial \varphi^i} - 
\nabla_\mu \left( \frac{\partial \mathcal L}{\partial (\partial_\mu \varphi^i)} 
\right) \right\} \delta \varphi^i + 
\nabla_\mu \left( \frac{\partial \mathcal L}{\partial (\partial_\mu \varphi^i)} 
\delta \varphi^i \right) \right].
\end{split}
\end{equation}
Thus, the field equation of motion for classical trajectories is given by:
\begin{equation} \label{fldeom} 
\frac{\partial \mathcal L}{\partial \varphi^i} - 
\nabla_\mu \left( \mathcal P^\mu_i \right) = 0 , \qquad  
\text{where } \quad \mathcal P^\mu_i :=
\frac{\partial \mathcal L}{\partial (\partial_\mu \varphi^i)}.
\end{equation}
To see how classical mechanics for single point particle is a field theory as 
well, setting 
$x^\mu \rightarrow \tau \; \Rightarrow \; d^n x \rightarrow d \tau$, and 
$g_{\mu \nu} (\bm x) \rightarrow g \rightarrow 1$ 
gives us the familiar action integral and the Euler-Lagrange 
equation for point particles. \smallskip 

When considering dynamics of $N+1$-dimensional scalar field space, we will 
essentially deal with the Polyakov form Lagrangian: 
\begin{equation} \label{ngeolag} 
\mathcal L = \frac12 g^{\mu \nu} (\bm x) 
h_{ij} (\bm \varphi) \partial_\mu \varphi^i \partial_\nu \varphi^j. 
\end{equation} 
and its Euler-Lagrange field equation given by \eqref{fldeom} as: 
\begin{equation} \label{fldgeo1} 
g^{\mu \nu} (\bm x) \left( \nabla_\mu \partial_\nu \varphi^i + 
\Omega^i_{jk} \partial_\mu \varphi^j \partial_\nu \varphi^k \right) = 0,
\end{equation} 
where the field space connection symbol $\Omega^i_{jk}$ in \eqref{fldgeo1} is 
\begin{equation} \label{conn} 
\Omega^i_{jk} = \frac{h^{im} (\bm \varphi)}2 
\left( \frac{\partial h_{mj} (\bm \varphi)}{\partial \varphi^k} + 
\frac{\partial h_{mk} (\bm \varphi)}{\partial \varphi^j} - 
\frac{\partial h_{jk} (\bm \varphi)}{\partial \varphi^m} \right). 
\end{equation}
Now we shall proceed to discuss how a conserved quantity 
may exist in scalar field theory as a result of a cyclic field.

\subsection{Cyclic fields} 
\label{sec:cyclic}

Even if the field Lagrangian $\mathcal L$ is independent of a cyclic field 
$\varphi^0 = \mathcal V$, the corresponding field momentum $\mathcal P^\mu_{\mathcal V}$ is not 
necessarily a conserved quantity, since according to \eqref{fldeom}, we have a 
covariant divergence equation. 
\begin{equation} \label{cyclic} 
\nabla_\mu \mathcal P^\mu_\mathcal V = \partial_\mu \mathcal P^\mu_\mathcal V + 
\Gamma^\mu_{\mu \rho} \mathcal P^\rho_\mathcal V = 
\frac{\partial \mathcal L}{\partial \mathcal V} = 0. 
\end{equation}
Now consider the following field Lagrangian based on the model \eqref{ngeolag} 
where $x^0 = y$ and $i, j = 1, . . . N$, and 
$\varphi^0 (\bm x) = \mathcal V (\bm x)$
\begin{equation} \label{lag1} 
\mathcal L = \frac12 g^{\mu \nu} (\bm x) 
\left[ h_{ij} (\bm \varphi) \partial_\mu \varphi^i \partial_\nu \varphi^j + 
2 h_{i0} (\bm \varphi) \partial_\mu \varphi^i \partial_\nu \mathcal V + 
h_{00} (\bm \varphi) \partial_\mu \mathcal V \partial_\nu \mathcal V \right]. 
\end{equation} 
Here in \eqref{lag1}, we can clearly see that if we set $\mathcal V (\bm x) = \mathcal V (y) 
\ \Rightarrow \ 
\partial_\mu \mathcal V = 0 \ \forall \ \mu \neq 0$
\[ \begin{split} 
\mathcal P^\mu_\mathcal V = 
\frac{\partial \mathcal L}{\partial (\partial_\mu \mathcal V)} &= 
g^{\mu \nu} (\bm x) \left[ h_{i0} (\bm \varphi) \partial_\nu \varphi^i + 
h_{00} (\bm \varphi) \partial_\nu \mathcal V \right]_{\mathcal V (\bm x) = 
\mathcal V (y)} \\ 
&= g^{\mu \nu} (\bm x) h_{i0} (\bm \varphi) \partial_\nu \varphi^i + 
g^{\mu 0} (\bm x) h_{00} (\bm \varphi) \partial_0 \mathcal V.
\end{split} \]
Thus, the only way to ensure that $\mathcal P^\mu_\mathcal V$ is restricted to 
only one non-zero component $\mathcal P^0_\mathcal V$ is to have vanishing 
cross terms in the base space and configuration space metric. ie.:
\begin{equation} \label{set1} 
g^{\mu 0} (\bm x) = 
h_{i0} (\bm \varphi) = 0, \ \mathcal V = \mathcal V (y) 
\qquad \Rightarrow \qquad 
\mathcal P^\mu_\mathcal V = 0 
\quad \forall \quad \mu \neq 0.
\end{equation}
Now, if $g_{\mu \nu} (\bm x)$ is independent of $x^0 = y$, 
\begin{equation} \label{set2} 
\partial_0 g_{\mu \nu} = 0 \qquad \Rightarrow \qquad \Gamma^\mu_{\mu 0} = 
\frac12 g^{\rho \sigma} \partial_0 g_{\rho \sigma} = 0.
\end{equation}
then upon applying \eqref{set1} and \eqref{set2} to \eqref{cyclic}, we shall 
have: 
\begin{equation} \label{const} 
\nabla_\mu \mathcal P^\mu_\mathcal V = \nabla_0 \mathcal P^0_\mathcal V 
\quad = \quad 
\partial_0 \mathcal P^0_\mathcal V = 
\frac{\partial \mathcal L}{\partial \mathcal V} = 0 \qquad \Rightarrow \qquad 
\mathcal P^0_\mathcal V = const.
\end{equation}
Thus, we have a field momentum that is a constant of motion. In summary, the 
conditions for this field momentum to be a constant of field dynamics are: 
\begin{enumerate}
\item the cyclic field $\mathcal V$ is dependent only on the cyclic co-ordinate $x^0 = y$, 
\item the metric of the co-ordinate space $g_{\mu \nu} (\bm x)$ is independent 
of $x^0 = y$, and 
\item cross-terms with cyclic co-ordinates and fields must vanish in the base and 
configuration space metrics (ie. $g^{\mu 0} (\bm x) = 0 \ \forall \ \mu \neq 0$, 
$h_{i0} (\bm \varphi) = 0$).
\end{enumerate}
We will next describe the generator of the field theory equivalent of Hamilton's 
equations of motion.

\subsection{A generator of field equations} 
\label{sec:fldgen}

Here, we shall introduce a function in fields and field momenta that generates 
the scalar field theory equivalent of Hamilton's equations of motion. In a 
similar manner to Legendre's method to define the Hamiltonian, the function 
shall be defined as:
\begin{equation} \label{gen} 
G (\varphi^i, \mathcal P^\mu_i) := \mathcal P^\mu_i \partial_\mu \varphi^i - 
\mathcal L.
\end{equation}
If we take the gradient of the generator \eqref{gen}, we can show that:
\begin{equation} \label{gengrad} 
\begin{split} 
\partial_\alpha G &= 
\frac{\partial G}{\partial \varphi^i} \partial_\alpha \varphi^i + 
\frac{\partial G}{\partial \mathcal P^\mu_i} \partial_\alpha \mathcal P^\mu_i \\ 
&= \partial_\mu \varphi^i \partial_\alpha \mathcal P^\mu_i + 
\mathcal P^\mu_i \partial_\alpha \partial_\mu \varphi^i - 
\left(\frac{\partial \mathcal L}{\partial \varphi^i} \partial_\alpha \varphi^i + 
\frac{\partial \mathcal L}{\partial (\partial_\mu \varphi^i)} \partial_\alpha 
\partial_\mu \varphi^i \right).
\end{split} 
\end{equation}
Applying \eqref{fldeom} to \eqref{gengrad}, we can see that:
$$\frac{\partial G}{\partial \varphi^i} \partial_\alpha \varphi^i + 
\frac{\partial G}{\partial \mathcal P^\mu_i} \partial_\alpha \mathcal P^\mu_i = 
\partial_\mu \varphi^i \partial_\alpha \mathcal P^\mu_i - 
\nabla_\mu \mathcal P^\mu_i \partial_\alpha \varphi^i,$$
where upon comparing the co-efficients of $\partial_\alpha \varphi^i$ and 
$\partial_\alpha \mathcal P^\mu_i$, we can see that:
\begin{equation} \label{fldhameom} 
\partial_\mu \varphi^i = \frac{\partial G}{\partial \mathcal P^\mu_i}, \qquad  
\nabla_\mu \mathcal P^\mu_i = - \frac{\partial G}{\partial \varphi^i}. 
\end{equation}
Like the Hamiltonian for point-particle mechanics, this generator is 
instrumental to the process of Eisenhart lifting the scalar field space.

\section{Eisenhart lift of scalar field space}
\label{lift}

So far, in \cite{fkp}, we have seen the Eisenhart-Riemannian lift of a field 
space by introducing vector fields. Here, we shall perform Eisenhart lift for 
field theory using a cyclic scalar field only. One must ensure that a cyclic 
co-ordinate wrt the co-ordinate space metric is available throughout the 
setup to ensure the existence of a conserved quantity that enables the lift 
via a cyclic field without disturbing the field equations \eqref{fldhameom}.

\subsection{Riemannian field lift }
\label{riemlift}

Let us suppose that we have a field Lagrangian \eqref{lag1} with the conditions 
\eqref{set1} and \eqref{set2} applied given by: 
\begin{equation} \label{liftfldlag1} 
\mathcal L = \frac12 g^{\mu \nu} (\bm x) h_{ij} (\bm \varphi) 
\partial_\mu \varphi^i \partial_\nu \varphi^j + 
\frac12 g^{00} (\bm x) h_{00} (\bm \varphi) 
\left( \partial_0 \mathcal V \right)^2,
\end{equation}
The generator \eqref{gen} in this case will be: 
\begin{equation} \label{liftgen2} 
G (\varphi^i, \mathcal P^\mu_i, \mathcal P^0_\mathcal V) = 
\frac12 g_{\mu \nu} (\bm x) h^{ij} (\bm \varphi) \mathcal P^\mu_i \mathcal P^\nu_j + 
\frac1{2 g^{00} (\bm x) h_{00} (\bm \varphi)} ( \mathcal P^0_{\mathcal V} )^2.
\end{equation} 
where the conserved field momentum is: 
$$\mathcal P^0_{\mathcal V} = g^{00} (\bm x) h_{00} (\bm \varphi) \; \partial_0 \mathcal V = q.$$ 
If we define a scalar field function: 
$$\Phi (\bm \varphi) := \frac{q^2}{2 g^{00} (\bm x) h_{00} (\bm \varphi)},$$ 
then the field equations \eqref{fldhameom} deduced from the lifted generator \eqref{liftgen2} are:
\begin{equation} \label{fldhameq1} 
\partial_\mu \varphi^i = \ \frac{\partial G}{\partial \mathcal P^\mu_i} = 
g_{\mu \nu} (\bm x) h^{ij} (\bm \varphi) \mathcal P^\nu_j, 
\end{equation}
which when applied to the lifted generator \eqref{liftgen2}, leads us to the field 
Lagrangian:
\begin{equation} \label{orglag1} 
\mathcal L = \frac12 g^{\mu \nu} (\bm x) h_{ij} (\bm \varphi) 
\partial_\mu \varphi^i \partial_\nu \varphi^j - \Phi (\bm \varphi),
\end{equation}
Thus, we can say that the Riemannian lift of the field Lagrangian \eqref{orglag1} is: 
\begin{equation} \label{liftlag1} 
\mathcal L = \frac12 g^{\mu \nu} (\bm x) h_{ij} (\bm \varphi) 
\partial_\mu \varphi^i \partial_\nu \varphi^j + 
\frac{q^2}{4 \Phi (\bm \varphi)} 
\left( \partial_0 \mathcal V \right)^2.
\end{equation}
\smallskip 

If a co-ordinate $x^0 = y$ cyclic wrt the metric $g_{\mu \nu} (\bm x)$ does 
not exist, we can create a new co-ordinate $y$ such that $g_{\mu 0} (\bm x) = 0 
\ , \ g_{00} (\bm x) = 1$, that only the new cyclic field created for the lift 
will depend on. This method will be called ``Double Lift'', since it expands the 
co-ordinate space alongside the field space. However, since the other fields are 
independent on $y$, it cannot be a generalization of point particle theory, 
since the fields in a point particle theory are dependent on the sole parameter.

\subsection{Lorentzian field lift}
\label{lorlift}

This time, we shall consider a field Lagrangian of the form: 
\begin{equation} \label{liftfldlag2} 
\mathcal L = \frac12 g^{\mu \nu} (\bm x) h_{ij} (\bm \varphi, \mathcal U) 
\partial_\mu \varphi^i \partial_\nu \varphi^j + 
g^{00} (\bm x) 
\left( \frac12 h_{\mathcal U \mathcal U} (\bm \varphi, \mathcal U) 
(\partial_0 \mathcal U)^2 + 
h_{\mathcal U \mathcal V} (\bm \varphi, \mathcal U) 
(\partial_0 \mathcal U) (\partial_0 \mathcal V) \right), 
\end{equation} 
where we have decided to set $g^{00} (\bm x) = 1, h_{\mathcal U \mathcal V} (\bm \varphi, \mathcal U) = 1$. 
The generator \eqref{gen} will be: 
$$G (\varphi^i, \mathcal U, \mathcal P^\mu_i, \mathcal P^0_\mathcal U) = 
\frac12 g_{\mu \nu} (\bm x) h^{ij} (\bm \varphi, \mathcal U) 
\mathcal P^\mu_i \mathcal P^\nu_j + 
\mathcal P^0_{\mathcal U} \mathcal P^0_{\mathcal V} - 
\frac12 h_{\mathcal U \mathcal U} (\bm \varphi, \mathcal U) (\mathcal P^0_{\mathcal V})^2.$$
We can write the conserved field momentum as a constant $\mathcal P^0_{\mathcal V} = q$ 
to write the generator \eqref{gen2} as: 
\begin{equation} \label{gen2} 
G (\varphi^i, \mathcal U, \mathcal P^\mu_i, \mathcal P^0_\mathcal U) = 
\frac12 g_{\mu \nu} (\bm x) h^{ij} (\bm \varphi, \mathcal U) 
\mathcal P^\mu_i \mathcal P^\nu_j + 
q \mathcal P^0_{\mathcal U} - 
\frac{q^2}2 h_{\mathcal U \mathcal U} (\bm \varphi, \mathcal U).
\end{equation}
Thus, the field equations generated according to \eqref{fldhameom} 
from the lifted generator \eqref{gen2} will be:
\begin{equation} \label{fldhameq2} 
\begin{split} 
\partial_\mu \varphi^i &= \ \frac{\partial G}{\partial \mathcal P^\mu_i} = 
g_{\mu \nu} (\bm x) h^{ij} (\bm \varphi, \mathcal U) \mathcal P^\nu_j, \\ 
\partial_0 \mathcal U &= \ \frac{\partial G}{\partial \mathcal P^0_{\mathcal U}} = q 
\qquad \Rightarrow \qquad 
\mathcal U (y) = q  \; y. 
\end{split}
\end{equation}
If we write 
$\frac{q^2}2 h_{\mathcal U \mathcal U} (\bm \varphi, \mathcal U) = - V (\bm \varphi, \mathcal U)$, 
the regular field Lagrangian deduced from the generator \eqref{gen2} will be
\begin{equation} \label{orglag2} 
L = \mathcal P^\mu_i \partial_\mu \varphi^i + 
\mathcal P^0_\mathcal U \partial_0 \mathcal U - 
G (\varphi^i, \mathcal U, \mathcal P^\mu_i, \mathcal P^0_\mathcal U) 
\quad = \quad \frac12 g^{\mu \nu} (\bm x) h_{ij} (\bm \varphi, \mathcal U) 
\partial_\mu \varphi^i \partial_\nu \varphi^j - 
V (\bm \varphi, \mathcal U). 
\end{equation} 
Thus, the field Lagrangian \eqref{liftfldlag2} deduced from lifting \eqref{orglag2} can be written as 
\begin{equation} \label{liftlag2} 
\mathcal L = \frac12 g^{\mu \nu} (\bm x) h_{ij} (\bm \varphi, \mathcal U) 
\partial_\mu \varphi^i \partial_\nu \varphi^j - 
\frac1{q^2} V (\bm \varphi, \mathcal U) (\partial_0 \mathcal U)^2 + 
(\partial_0 \mathcal U) (\partial_0 \mathcal V). 
\end{equation} 
Other applications of the Eisenhart-Duval lift to branes are discussed in \cite{Hassaine2001}
where the authors study the relativistic generalization of the Chaplygin gas in Duval's 
Kaluza-Klein framework, and in \cite{Gibbons2020},  where Gibbons applies the Eisenhart-Duval 
lift to a flat Lorentzian spacetime with two times.

\section{Randers-Finsler formulation for Nambu-Goto action} 
\label{nambu}

Randers \cite{Randers} performed the first significant application of Finslerian geometry to physics, 
describing the action in terms of the charged massive particle's worldline and its interaction with 
gauge fields to account for the influence of curvature of spacetime and gauge fields in its motion, the 
related mechanics of which has been discussed in detail in \cite{chanda}. The Finslerian connection 
for massless particles was discussed by Duval \cite{Duval2007} and Elbistan, Zhang, Dimakis, 
Gibbons,  and Horvathy in \cite{Elbistan2020}.  In the former, Duval formulates a general theory 
of geometric optics for spinning light rays on a Finsler manifold, effectively deducing Finslerian 
spinoptics, while the authors in the latter study the free motion of a massive particle in a Finslerian 
deformation of a plane gravitational wave. \smallskip

So far, we have discussed the Lagrangian for a scalar field theory written in quadratic form 
of a regular scalar field Lagrangian. However, they often have a geometric origin, such as the 
world volume spanned in the scalar field space. One such example is the Nambu-Goto action 
discussed in string theory where the action is directly proportional to the total surface area 
of a worldsheet swept out in spacetime. \bigskip 

\noindent
If the action be written as the integral of the volume spanned in configuration space:
\begin{align} 
S &= \int d^n x L = \int d^n x \sqrt{M}, \nonumber \\ 
\label{detact} 
\text{where } 
M &= Det(M_{\mu \nu}) 
, \quad 
M_{\mu \nu} = h_{ij} (\bm \varphi) 
\frac{\partial \varphi^i}{\partial x^\mu} 
\frac{\partial \varphi^j}{\partial x^\nu}, 
\end{align} 
then we can deduce the scalar field momentum from \eqref{detact} to be: 
\begin{equation} \label{strmom} 
\mathcal P^\mu_i = \frac{\partial L \ }{\partial (\partial_\mu \varphi^i)} = 
\frac{\partial (\sqrt{M})}{\partial (\partial_\mu \varphi^i)} = 
\frac12 \sqrt{M} \; M^{\alpha \beta} 
\frac{\partial M_{\alpha \beta}}{\partial (\partial_\mu \varphi^i)} = 
\sqrt{M} \; h_{ij} (\bm \varphi) M^{\mu \nu} \frac{\partial \varphi^j}{\partial x^\nu}
\end{equation} 
Here, we can see from applying \eqref{strmom} that a rule analogous to the Maupertuis 
principle for point particle mechanics exists for such geometric actions. 
\begin{equation} \label{fldmaup} 
\mathcal P^\mu_i \partial_\mu \varphi^i = 
\sqrt{M} \; h_{ij} (\bm \varphi) M^{\mu \nu} 
\frac{\partial \varphi^j}{\partial x^\nu} 
\frac{\partial \varphi^i}{\partial x^\mu} = 
\sqrt{M} \; h_{ij} (\bm \varphi) M^{\mu \nu} M_{\mu \nu} = 
n \sqrt{M}. 
\end{equation} 
In case of strings, we will have $n = 2$, and in case of a point particle $n = 1$ for 
which the familiar Maupertuis principle discussed in classical mechanics, easily evident for 
Randers-Finsler geodesics \cite{chanda,cgg1,cggmw} re-emerges. One can easily see 
from \eqref{fldmaup} that for world volumes, the trace of the energy momentum tensor 
shall vanish, similar to the overall Hamiltonian for Randers-Finsler geodesics as shown 
in \cite{chanda,cgg1}. 
\begin{equation} \label{emtensor} 
T^\mu_\mu = 
\frac{\partial L \ }{\partial (\partial_\mu \varphi^i)} 
\partial_\mu \varphi^i - \delta^\mu_\mu L = 
\mathcal P^\mu_i \partial_\mu \varphi^i - n L  = 0.
\end{equation}
If any one of the fields $\varphi^0 = \mathcal V$ is cyclic, and satisfies all the conditions 
mentioned for its conjugate field momentum $\mathcal P^0_{\mathcal V}$ to be conserved 
as described in Sect.~\ref{sec:cyclic}, then we will have from \eqref{fldmaup}: 
\begin{equation} \label{addit} 
L = \sqrt{M} = \frac1n \sum_{i \neq 0} \mathcal P^\mu_i \partial_\mu \varphi^i + 
\partial_\mu \left( \frac1n \mathcal P^\mu_{\mathcal V} \mathcal V \right) = 
\frac1n \sum_{i \neq 0} \mathcal P^\mu_i \partial_\mu \varphi^i + 
\partial_\mu f^\mu. 
\end{equation}
As we know, the last term for a field theory being a total divergence term 
is an additive freedom factor that can be omitted from the Lagrangian 
\eqref{addit} to formulate a Jacobi Lagrangian. 
\begin{equation} \label{jaclag} 
L_J = \frac1n \sum_{i \neq 0} \mathcal P^\mu_i \partial_\mu \varphi^i.
\end{equation} 
However, there is no constraint for a geometric world volume action in curved 
space, except for a point particle, so we are unable to determine a Jacobi 
metric as done in \cite{chanda,cggmw}. This is evident if one uses 
\eqref{strmom} to write a matrix: 
\begin{equation} \label{mommat} 
C^{\mu \nu} = 
h^{ij} (\bm \varphi) \mathcal P^\mu_i \mathcal P^\nu_j = 
M M^{\mu \nu} 
\qquad \Rightarrow \qquad 
C_{\mu \nu} = M^{-1} M_{\mu \nu},
\end{equation} 
\begin{equation} \label{momdet} 
C = Det(C^{\mu \nu}) = M^n Det(M^{\mu \nu}) = M^{n-1}.
\end{equation}
Thus, we can see that only for a point particle $n = 1$ we will have the constraint, 
described in \cite{chanda,cggmw}. 
$$C_{n = 1} = 1.$$
However, for the Nambu-Goto action of string, we can formulate a generator of 
a field equation from \eqref{momdet} using \eqref{strmom} and \eqref{mommat}  
\begin{equation} \label{geomgen} 
G = \sqrt{C_{n = 2}} 
, \qquad \text{where } \ 
C_{n = 2} = M 
\end{equation} 
\begin{align} 
\frac{\partial G}{\partial \mathcal P^\mu_i} &= 
\frac12 \sqrt{C_{n = 2}} \; C_{\alpha \beta} 
\frac{\partial C^{\alpha \beta}}{\partial \mathcal P^\mu_i} = 
\sqrt{M} \; M^{-1} M_{\mu \beta} 
h^{ij} (\bm \varphi) \mathcal P^\beta_j = 
\frac{\partial \varphi^i}{\partial x^\mu} \nonumber \\ 
\label{geneom} 
\frac{\partial G}{\partial \varphi^i} &= 
\frac12 \sqrt{C_{n = 2}} \; C_{\alpha \beta} 
\frac{\partial C^{\alpha \beta}}{\partial \varphi^i} = 
\Omega^j_{ik} \mathcal P^\beta_j \frac{\partial \varphi^k}{\partial x^\beta} = 
\nabla_\mu \mathcal P^\mu_i,
\end{align}
where in the 2nd equation of \eqref{geneom}, $\Omega^j_{ik}$ is the connection 
on the scalar field space introduced in \eqref{conn}, and the equation 
$$\nabla_\mu \mathcal P^\mu_i = 
\Omega^j_{ik} \mathcal P^\beta_j \frac{\partial \varphi^k}{\partial x^\beta},$$
can easily be verified by applying the Lagrangian $L = \sqrt{M}$ (all constant factors 
adjusted to 1) for Nambu-Goto action the Euler-Lagrange field equation \eqref{fldeom}. 
Such a generator can be formulated only for point particle and the Nambu-Goto string. 
Furthermore, the Randers-Finsler form for the Nambu-Goto action can intuitively be 
written as: 
\begin{equation} \label{ngrf} 
S = \int d \tau d \sigma \left( \sqrt{M} + 
\rho(\bm \sigma) A_i (\bm \varphi) \partial_\tau \varphi^i \right). 
\end{equation}
where $\rho (\sigma)$ is the density along the string, and $\bm A (\bm \varphi)$ 
are the gauge fields in scalar field space. In this case, we can define a gauge covariant 
field momentum 
\begin{equation} \label{gcov} 
\Pi^\tau_i = \mathcal P^\tau_i - \rho(\bm \sigma) A_i (\bm \varphi) = 
\sqrt{M} \; h_{ij} (\bm \varphi) M^{\tau \nu} \frac{\partial \varphi^j}{\partial x^\nu},
\end{equation}
which will substitute the canonical field momentum $\mathcal P^\tau_i$ in the 
formulation of the generator of equations for a string introduced previously in 
\eqref{geomgen}, as done in \cite{chanda,cggmw}.

\section{Conclusion and Discussion}

We started by reviewing scalar field theory, showing how a conserved quantity 
can exist in scalar field theory, and introduced a generator of the field theory 
equivalent of Hamilton's equations of motion. Such conditions are not necessary 
for point particle, where non-zero cross terms in the field space metric 
are allowed. \smallskip

We then showed how to perform both, the Riemannian and Lorentzian versions 
of the Eisenhart lifts, for n-dimensional scalar field theories, using the Lagrangian 
of the Polyakov form. While it is restricted by specific conditions 
compared to that described in \cite{fkp}, it is more similar and comparable to the 
procedure for point particle mechanics, and can be considered the simplest 
and most direct extension of the geometric lifting procedure to scalar field theory. 
\smallskip  

If a cyclic co-ordinate that the metric is independent of is unavailable, the co-ordinate 
space could also be expanded to include one, describing a ``Double Lift''. However, a 
``Double Lifted'' field theory cannot be considered a generalization of point particle 
theory since the fields are independent of the cyclic co-ordinate. \smallskip 

Finally, we can see that many mechanical principles associated with point particle, 
such as Maupertuis principle can be generalized to world volume actions, for which 
one setting gives the Nambu-Goto action. Furthermore, a generator of equations 
analogous to a Hamiltonian can be described only for strings and point particles, 
where in the latter case, it is constrained to a constant value.

\section*{Acknowledgement}

We wish to acknowledge G. W. Gibbons, M. Cariglia, and Joydeep Chakravarty for 
various discussions and support, P. Horvathy, K. Morand and A. Galajinsky for 
supportive comments, and P. Mukhopadhay for introducing us to scalar field theory, 
all of which were instrumental in the preparation of this article.

\end{document}